# APHRODITE: an Anomaly-based Architecture for False Positive Reduction[*]

Damiano Bolzoni, Sandro Etalle

University of Twente,
P.O. Box 2100, 7500 AE Enschede, The Netherlands
{damiano.bolzoni, sandro.etalle}@utwente.nl

**Abstract.** We present APHRODITE, an architecture designed to reduce false positives in network intrusion detection systems. APHRODITE works by detecting *anomalies* in the *output traffic*, and by correlating them with the alerts raised by the NIDS working on the input traffic. Benchmarks show a substantial reduction of false positives and that APHRODITE is effective also after a "quick setup", i.e. in the realistic case in which it has not been "trained" and set up optimally.
**Keywords:** Intrusion Detection, False Positives

## 1 Introduction

Network intrusion detection systems (NIDSs) are considered an effective second line of defense against network-based attacks directed at computer systems [1,2], and – due to the increasing severity and likelihood of such attacks – are employed in almost all large-scale IT infrastructures [3].

The Achille's heel of NIDSs lies in the large number of *false positives* (i.e., false attacks) that occur: practitioners [4,5] as well as researchers [6,7,8] observe that it is common for a NIDS to raise thousands of mostly false alerts per day. Manganaris et al. [4] were able to collect more than 15,000 alerts per day per sensor during a monitoring period of just one month. Julisch [9] states that up to 99% of total alarms can be false alarms. Indeed, a high rate of false alarms is – also according to Axelsson [6] – the limiting factor for the performance of an intrusion detection system. False alarms often cause an overload for IT personnel [4], who must verify every single alert to prevent or block possible loss of data confidentiality, integrity and availability (CIA). The manual identification of true positives amongst this flood of alarms it is not only labor intensive but also error prone [10].

False positives are a universal problem as they affect both *signature-based* intrusion detection systems and *anomaly-based* systems [11]. Finally, a high false positive rate can even be *exploited* by attackers to overload IT personnel, thereby lowering the defenses of the IT infrastructure.

[*] This research is supported by the research program Sentinels (http://www.sentinels.nl). Sentinels is being financed by Technology Foundation STW, the Netherlands Organization for Scientific Research (NWO), and the Dutch Ministry of Economic Affairs.

*Contribution of this paper* Our thesis is that one of the main reasons why NIDSs show a high false positive rate is that they do not *correlate* input with output traffic: by observing the output determined by the alert-raising input traffic, one is capable to reduce the number of false positives in an effective manner.

To demonstrate this, we have developed *APHRODITE* (Architecture for false Positives Reduction): an innovative architecture for reducing the false positive rate of *any* NIDS (be it signature-based or anomaly-based). APHRODITE consists of an output anomaly detector (OAD) and a *correlation engine*; in addition, APHRODITE assumes the presence of a NIDS on the input of the system (see Figure 1).

We have benchmarked APHRODITE in combination with the signature-based NIDS Snort [12,13], as well as APHRODITE in combination with the anomaly-based NIDS POSEIDON. We have carried out the benchmarks both on the common DARPA 1999 data set [14] as well as on a private data set. In 6 out of 7 cases, our benchmarks show a reduction of false positives between 50% and 100%, which is better than the only leading competitor [15] providing benchmarks on a public data set.

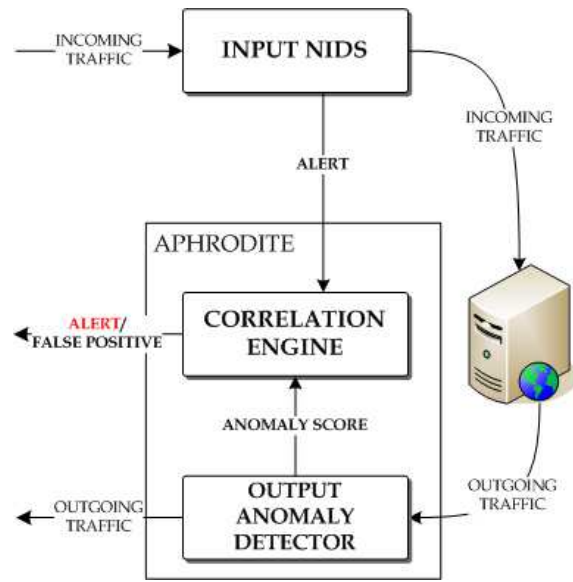

**Fig. 1.** APHRODITE architecture

*Architecture* The main idea of our approach is simple: a successful attack often causes an *anomaly* in the output of the system. For instance, a successful SQL injection against a web-based system would typically cause the system to output e.g. SQL tables rather than the usual web pages. Taking advantage of this,

APHRODITE works as follows: when the NIDS raises an alert, the correlation engine checks whether the communication that raised this alert also causes an anomaly in the output (detected by the OAD). If this is the case, the alert is considered a *true positive* and APHRODITE forwards it to the IT professionals, otherwise, it is discarded as a *false positive*. (There are various exceptions to this behaviour, taking into account e.g., the possible *absence* of output or the quality/quantity of alerts raised by the NIDS, we explain this in Section 3.)

*Quick setup* Our benchmarks show that APHRODITE is effective also after a "quick setup", without an optimal training and by using a simple heuristics for setting the threshold. This is particularly relevant because anomaly detection systems (like our OAD) are often regarded as systems whose deployment is rather labor-intensive. This is because they need to be trained with a considerable amount (gigabytes) of data, which should be as clean as possible: the training data set should be representative of attack-free traffic. To carry out such a training, in principle the IT professional should clean up the data set by inspecting it and by manually removing the spurious traffic, a procedure which is time consuming. Moreover, the anomaly detector should be re-trained each time that changes in the system denote a change in the traffic type. In addition, anomaly-detection systems require the IT professional to spend time to set the *threshold* (see also Section 2). In our case we show that APHRODITE is reasonably accurate and complete also when it is trained with a data set that was not cleaned up manually, and when the threshold is set using simple heuristics. This makes APHRODITE attractive for real-life situations, in which the IT professionals want to have a tool helping them to isolate true positives without requiring a troublesome set up.

*Structure of the paper* This paper is organized as follows: in Section 2 we introduce intrusion detection systems and the problems related to false positive. Section 3 reports the system architecture and its properties. In Section 4 we report the results of our benchmarks. In Section 5 we discuss other related work. Finally, in Section 6 we draw our conclusions and set the course for further developments. In the Appendix we report the pseudo-code of APHRODITE.

## 2 Preliminaries

In this Section, we introduce the concepts used in the rest of the paper and we explain in more detail than in the introduction how false positives arise and the harm they can do to a system. Those who are familiar with signature-based and anomaly-based intrusion detection systems may skip this part.

There exist two different sorts of network intrusion detection systems: signature-based and anomaly-based; both types are affected by a high false positive rate.

### 2.1 Signature-Based Systems

Signature-based systems (e.g. Snort [12,13]) are based on pattern-recognition techniques: the NIDS contains a database of *signatures* of known attacks and tries to match these signatures against the analyzed data. When a match is found, an alarm is raised. A signature must be developed off-line, and then loaded in the database before the system can begin to detect a particular intrusion. One of the disadvantages of signature-based NIDS is that they can detect only known attacks: *all* new attacks will go unnoticed until the system is updated, creating a window of opportunity for attackers (and affecting the NIDS completeness and accuracy [16,2]). Although this is considered acceptable for detecting attacks to e.g., the OS, it makes them less suitable for protecting web-based services, because of their ad-hoc and dynamic nature.

**False Positives in Signature-Based Systems** Signature-based systems raise an alarm each time that traffic matches one of the signatures loaded into the system. For example: the *path traversal attack* allows access to files, directories, and commands that reside outside the (given) web document root directory. The most elementary path traversal attack uses the `../` character sequence to alter the resource location requested in the URL. Variations include valid and invalid Unicode-encoding ("`..%u2216`" or "`..%c0%af`"), URL encoded characters ("`%2e%2e%2f`"), and double URL encoding ("`..%255c`") of the backslash character (excerpted from *WASC Threat Classification* [17]).

To detect these attacks, Snort, with the default configuration (like any signature-based intrusion detection system), raises an alarm each time that it identifies the pattern `../` in the incoming traffic. Unfortunately, this pattern is also often present in legal traffic, and this causes Snort to raise a high number of false alarms. These false alarms can be avoided by deactivating this rule, but this on the other hand prevents the NIDS from finding this sort of attacks.

*Tuning Signature-Based Systems* A common way to reduce false positives in signature-based systems is by deactivating the signatures relative to vulnerabilities that are not present in the given environment: many signatures can be disabled because the monitored services are not exposed to a certain vulnerability or the vulnerability itself affects only certain OS platforms. This configuration phase of signature-based systems is also known as *tuning* and it is explicitly required when deploying NIDSs with several sensors in complex environments.

Tuning requires a thorough analysis of the environment by qualified IT personnel and the tuning must be kept up-to-date to keep up with the changes in the system: new vulnerabilities are discovered every day, new signatures are released regularly, and systems may be patched, thereby removing vulnerabilities.

### 2.2 Anomaly-Based Systems

Anomaly-based systems (ABS), unlike signature-based systems, first build a *statistical model* describing the normal network traffic, then flag any behaviour that significantly deviates from the model as an attack. This is achieved by implementing a *distance* function and setting a *threshold* value: when the distance between the input sample and the model exceeds the threshold, an alert is raised.

The main advantage of ABSs is that they can detect zero-day attacks: novel attacks can be detected as soon as they take place. The disadvantage is that an ABS requires an extensive model building phase: a significant amount of (largely attack-free) data must be analyzed to build accurate models of legal behaviour.

**False Positives in Anomaly-Based Systems** The value of the threshold has a direct influence on both false negatives and false positives rates [18]: a low threshold yields a high number of alarms, and therefore a low false negative rate, but a high false positive rate. On the other hand a high threshold yields a low number of alarms in general (therefore a high number of false negatives, but a low number of false positives). As a matter of fact, the high false positive rate is generally cited as one of the main disadvantages of anomaly-based systems.

*Tuning Anomaly-Based Systems* The most commonly used tuning procedure for ABS is finding an optimal threshold value, i.e., the best compromise between a low number of false negatives and a low (or acceptable) number of false positives. This is typically carried out manually by trained IT personnel: different improving steps can be necessary to obtain a good balance between detection rate and false positive rate.

### 2.3 Exploiting false positives

To conclude this section, we note that the presence of a NIDS with a relatively high number of false positives can also be *exploited* by an attacker.

Both signature-based and anomaly-based systems can be fooled to raise thousands of alerts: in the former case, the attacker can force an alert by inserting a known attack sequence randomly in the input (e.g. inside a field of a web page form), even if it is not aimed to exploit any vulnerability; in the latter case, the attacker can send an input that is substantially different from the data analyzed during the training phase (e.g. a long stream with the same character, which obviously was never observed before).

This can be exploited by an attacker, e.g., to perpetrate a Denial of Service attack against IT personnel [4] or to hide a real attack.

## 3 Architecture

The main idea of our approach is the following: a successful attack (incident) [3] to a system (e.g., a web service) usually produces an anomaly in the *output* of the system. On the other hand, if something in the input of the system causes the NIDS to raise an alarm but does not cause the system to produce an *unusual* output, then this alarm is likely to be a false positive.

*Example 1.* SQL Injection is a technique that exploits vulnerabilities of (web-based) applications which are interfaced to an SQL database: if the application does not sanitize potentially harmful characters first [17], an intruder can *inject* an SQL query in the database, and force the database to output sensitive data (e.g. user passwords and personal details) from database tables, without being authorized. SQL Injections are considered a serious threat and are constantly listed in the "Top Ten Most Critical Web Application Security Vulnerabilities" [19] by "The Open Web Application Security Project".

For instance, the following HTTP request is actually a well-known attack [20] against the Content Management System (CMS) PostNuke [21] that can be used to get hold of the user passwords:

```
http://[target]/[postnuke_dir]/modules.php?op=modload
   &name=Messages&file=readpmsg&start=
   0%20UNION%20SELECT%20pn_uname,null,pn_uname,pn_pass,pn_p
```

When such an attack is carried out successfully, the output (a database table) is significantly different from the HTML page usually rendered. This is exploited by our system to distinguish false positive from true positives.

*APHRODITE* works by detecting anomalies in the *output* of the system and by correlating them to the alerts raised by the NIDS monitoring the input of the system. We want to stress that the NIDS monitoring the input is essential for the functioning of APHRODITE, but it is not part of APHRODITE's architecture: as a matter of fact

APHRODITE can work together with *any kind of* NIDS (be it anomaly-based or signature-based).

The central component of APHRODITE is the *Output Anomaly Detector* (OAD), which is an anomaly-based NIDS placed on the output channel: the OAD refers to a statistical model describing the normal output of the system, and flags any behaviour that significantly deviates from the norm as the result of a possible attack. To connect the input to the output NIDS, the correlation engine is implemented with a stateful-inspection mechanism [22] to track and correlate input and output data belonging to the same communication.

APHRODITE works as follows (see Figure 1): the external NIDS monitors the input data while, simultaneously, the OAD analyzes the response of network services; when the input NIDS raises an alarm, it warns the *correlation engine* (CE), indicating the endpoint information (i.e. source and destination IP addresses, source and destination TCP ports) of the packet that raised the alert. At this moment, the alert is not considered an incident yet (it is a *pre-alarm*) and is not forwarded immediately to the IT specialists yet. Next, the correlation engine marks the communication relative to the given endpoints as suspicious and waits for the output of the OAD: if the OAD detects an anomaly in the outgoing traffic related to the tracked communication, then the system considers the alert as an *incident* (i.e. a true positive) and the alert is forwarded to the IT specialists for further handling and countermeasures reactions, otherwise it is considered a false positive.

What we just described is the most common behaviour; nevertheless there exist important exceptions to it. Once an alarm has been raised by the input NIDS, the following exceptions can take place:

1. **Missing output response**. If the OAD does not detect *any* output related to the pre-alarm raised by the NIDS, then the pre-alarm is considered a true incident and is forwarded to the IT specialist.
   This is because the pre-alarm could belong to a denial of service attack against a certain network service (preventing normal functioning or causing a complete stop), leading to a situation of missing response.
   Therefore, the absence of an output should be considered an anomaly in the normal data flow and must be handled accordingly by the correlation engine. To this end, we need to set appropriate application-dependent *time-outs* (commonly a data exchange between peers take place in a short time): after their expiration the communication can be marked as anomalous.
2. **Alarm magnitude**. When the NIDS monitoring the input is *anomaly-based* (as opposed to signature-based), then when it raises an alert it can also indicate the *magnitude* of the alert: anomaly-based NIDS compare the traffic to a statistical model of the traffic, and raise an alert when the input sample exceeds a give *threshold*. Here, by alarm magnitude we indicate the distance between the alarm-raising packet and the threshold. The higher the alarm magnitude, the more anomalous is the packet, and therefore the more likely that it indicates a true incident.
   In APHRODITE when the magnitude is higher than a given value the alert raised by the NIDS is considered an incident, even if the OAD has not detected any anomaly in the output.
3. **Number of alarm-raising packets**. APHRODITE takes into account the number of anomalies regarding output traffic related to a single endpoint (e.g. in the past and in the current communications) in a given time frame: this parameter becomes particularly interesting when the input NIDS is anomaly-based *and packet-oriented*, which can mark as anomalous a number of packets belonging to the same communication.

### 3.1 The OAD

As we mentioned before, the OAD is basically a payload-based anomaly-based intrusion detection system, which monitors the output of a system rather than the input of it. Specifically, the OAD has the same structure of POSEIDON [23], i.e. it is a payload-based two-tier NIDS, in which the first tier consists of a Self-Organizing Map (SOM), and is used exclusively to classify payload data; the second tier (the actual analyzer) consists of a slight modification of the well-known PAYL system [24]. Actually, for the OAD we could have used any anomaly-based payload-based NIDS, we chose POSEIDON because we are familiar with it and because it gives better results than leading competitors[23].

The pseudo-code of APHRODITE (the OAD and the correlator) is reported in Appendix A. The fact that the OAD is anomaly-based (rather than signature-based) allows it to adapt to the specific network environment/service, and to work in an unsupervised manner (at least, after the tuning). The disadvantage is that the OAD needs an extensive (though unsupervised) *training* phase: a significant amount of data is needed to build an accurate statistical model of the legal behaviour. During the training phase, there are some parameters that can affect the completeness and accuracy of the OAD, namely:

- **Duration of the training phase**. The duration of the training (thus, the number of samples used to train the system) directly affects the quality of the model that will be used in the detection phase: a too short training phase could lead to a (too) coarse data classification, which – in the detection phase – translates into flagging legitimate traffic too often as anomalous. The IT specialist can compensate for a too short training phase by increasing the *threshold* (see below), but this has the disadvantage that the OAD will classify (more) traffic resulting from attacks as legitimate.
- **Quality of the training phase**. Also the quality of the samples used during the training influences the quality of the models. The samples should be representative of normal behaviour and be as *attack-free* as possible, otherwise the OAD might classify data resulting from an attack as legitimate output traffic. In general, the more attack-free the data is, the more accurate the model will be. It is out of the scope of this work to detect and mitigate this problem but it is worth mentioning that we have obtained good results also using a private data set which was not made attack-free by hand (see Section 4 for further details).
- **Threshold**. As we said before, setting the threshold is a delicate task that could require different improving steps to reach a good balance between true positives rate and false positives rate.

## 4 Experiments and results

To validate our architecture, we have benchmarked APHRODITE in combination with the signature-based NIDS Snort [12,13] as well as APHRODITE in combination with the anomaly-based NIDS POSEIDON [23]. To do this, we have employed two different data sets: first, we have used the DARPA 1999 data set [14]: despite criticism [25,26] this is a standard data for benchmarking NIDS (e.g. [24,15]) and has the advantage that it allows one to duplicate experiments and to compare different NIDS directly. Secondly, we have benchmarked the system using a private data set.

**Tests with the DARPA 1999 data set** The testing environment of DARPA 1999 data set contains several internal hosts that have been attacked by both external and internal attackers. Moreover, hosts inside the local area network are able to conduct attacks against external hosts. In our tests, we focus on FTP, Telnet, SMTP and HTTP protocols. There are two reasons for this: first only these protocols gave us a sufficient number of samples we needed to train the OAD, and secondly only these protocols allowed to compare our architecture with POSEIDON, that has been benchmarked following the same procedures (because of the large sample set available only for these protocols). Other restrictions have been applied to make the comparison: we consider only inbound and outbound TCP packets that belong to attack connections against hosts of the network 172.016.0.0/16.

| Protocol | | Snort | Snort + APHRODITE | POSEIDON | POSEIDON + APHRODITE |
|---|---|---|---|---|---|
| HTTP | DR | 59,9% | 59,9% | 100% | 100% |
|  | FP | 599 (0,069%) | 5 (0,00057%) | 15 (0,0018%) | 0 (0,0%) |
| FTP | DR | 31,75% | 31,75% | 100% | 100% |
|  | FP | 875 (3,17%) | 317 (1,14%) | 3303 (11,31%) | 373 (1,35%) |
| Telnet | DR | 26,83% | 26,83% | 95,12% | 95,12% |
|  | FP | 391 (0,041%) | 6 (0,00063%) | 63776 (6,72%) | 56885 (5,99%) |
| SMTP | DR | 13,3% | - | 100% | 100% |
|  | FP | 0 (0,0%) | - | 6476 (3,69%) | 2797 (1,59%) |

**Table 1.** Comparison between Snort stand-alone, Snort in combination with APHRODITE, POSEIDON stand-alone and POSEIDON in combination with APHRODITE using the DARPA 1999 data set: DR stands for detection rate (attack instances), while FP is the false positive rate (packets); APHRODITE reduces FP by more than 50% most of the times, being close to zero in 3 tests, without affecting the detection rate.

We trained the OAD of APHRODITE with the data of weeks 1 and 3 (attack free): for each different protocol we have used a different OAD instance. Afterward, we tested APHRODITE together with both POSEIDON and Snort to on the traffic of weeks 4 and 5. The authors of DARPA provide a table containing all the attack instances, allowing one to distinguish between false positives and true positives attacks.

Table 1 reports a comparison of the detection rate and false positives rate of Snort stand-alone (first column), Snort in combination with APHRODITE (second column), POSEIDON stand-alone (third column) and POSEIDON in combination with APHRODITE (fourth column). In both cases, APHRODITE achieves a substantial improvement on the stand-alone system without affecting the detection rate nor introducing false negatives. APHRODITE has not be applied to SMTP traffic in combination with Snort because in this case Snort raises no false positives.

**Tests with a private data set** To complete our validation, and to see how the system behaves when trained with a data set that is not made attack-free, we considered a second (private) data set we have collected at the University of Twente: this is *data set B*. Data was collected on a public network for 5 consecutive working days (24 hours per day), logging only TCP traffic directed to (and originating from) a heavy-loaded web server (about 10 Gigabyte of total data per day). This web server hosts several official department web sites and personal web pages of students and research staff: thus, the traffic contains diverse data such as static and dynamically generated HTML pages and, especially in the output traffic, common format documents (e.g. PDF) as well as raw binary data (e.g. software executables). We did not inject any artificial attack.

We focus on HTTP traffic because nowadays Internet attacks are mainly directed to web-based services: Symantec Corporation reports that in year the 2005 web-based services have been the third most attacked service (ranked by TCP port), and in the second-half of the year 69% of total discovered vulnerabilities apply to web applications [27]. We classified alerts manually and detected 33 attack instances in 59288 input packets: most of the attacks are XSS (Cross-site Scripting) [17] and SQL Injection attacks. Table 2 summarizes the results we obtained.

We could not compare APHRODITE in combination with Snort on this second data set for the simple reason that Snort did not find any attack to the system (Snort raised only false alarms): by setting a high output threshold in APHRODITE we could have easily removed *all* the false positives, but this would have given no indication of the completeness and accuracy of APHRODITE.

| Protocol | | POSEIDON | POSEIDON + APHRODITE |
|---|---|---|---|
| HTTP | DR | 100% | 100% |
| | FP | 1683 (2,83%) | 774 (1,30%) |

**Table 2.** Comparison between POSEIDON stand-alone and POSEIDON in combination with APHRODITE using data set B; DR stands for detection rate (attack instances), while FP is the false positive rate (packets); APHRODITE reduces FP by more than 50% without affecting the detection rate.

*Quick setup* To train the anomaly-detection engines of both POSEIDON and the OAD on the data set B, we simply used a snapshot of the data collected during *working hours* (approximately 3 hours, 1,8 Gigabyte of data): it is widely acknowledged that attackers prefer to conduct malicious activity during non-working hours, when the system is usually less monitored by IT personnel. The chosen training data set has not been pre-processed and made attack-free: thus, it is possible that some malicious activities have been processed during the model building phase. For the same reason, we randomly choose a nightly snapshot (approximately 8 hours, 1,8 Gigabyte of data) to benchmark POSEIDON stand-alone against POSEIDON in combination with APHRODITE. Moreover, we setup the threshold following the simple heuristics discussed below.

This way of training the anomaly detection engines of POSEIDON and the OAD is not optimal (as we remarked before, the training set is supposed to be attack-free), but this allowed us to check the completeness and accuracy of our system in a fairly realistic situation, and to show that APHRODITE is useful for reducing false positives also in those cases in which the IT professional wants to carry out a *quick setup* and does not want to spend too much time cleaning up the training set and setting an optimal threshold, applying several enhancing steps.

**Setting the threshold** In Section 2.2 we introduce the fact that – in anomaly-based systems – completeness and accuracy are intrinsically related and they are heavily influenced by the *threshold value*. Here, we call *completeness* the ratio $TP/(TP+FN)$ and *accuracy* the ratio $TP/(TP+FP)$, where $TP$ is the number of true positives, $FN$ is the number of false negatives and $FP$ is the number of false positives raised during the benchmarks. Our experiments show that setting the threshold at $\frac{3t_{max}}{4}$, usually yields reasonably good results; here $t_{max}$ is the maximum distance between the analyzed data and the model observed during the training phase.

Tables 1 and 2 report the best false positive rate we have measured during our benchmarks without affecting the detection rate achieved during the stand-alone session. Figure 2 shows more accurately what happens to the accuracy and completeness of POSEIDON and POSEIDON in combination with APHRODITE when we modify the threshold of POSEIDON stand-alone (broken line) and when we modify the threshold of APHRODITE after having fixed that of POSEIDON to the best value (unbroken line). Here, we concentrate on the Telnet and SMTP protocol data of the DARPA 1999 data set and HTTP protocol data from data set B (these protocols presented the highest $FP$ rate, allowing more accurate measurements).

## 5 Related work

In this section we present related work. The problem of reducing false positive has been addressed using two different kind of approaches: on the one hand we have techniques for *identifying true positives*, and on the other hand we have techniques for *identifying false positives*.

The main difference between our work and the papers described below (with the exception of Qiao and Weixin's [28] – see below) consists of the fact that we take into account the output traffic of the system.

### 5.1 Identifying true positives

Ning et al. develop a model [29] and an intrusion alert correlator [30] to help human analysts during the alert verification phase. Their work is based on the observation that most incidents consist of several related stages, with the early stages preparing for the later ones. The authors introduce the concept of *prerequisite of an attack*: which is defined as the necessary condition for the attack to be successful. Furthermore, logical formulas are used to describe relationships between different attack stages, and hyper-alert correlation graphs are employed to represent correlated alerts in an intuitive way. However, this correlation technique is ineffective when attackers use a different source at each attack step.

412Ning and Cui [30] demonstrate the effectiveness of this approach when applied on a small data set: in [31,32] the same authors present other utilities they developed to facilitate the analysis of large sets of correlated alerts, and report some benchmarks employing network traffic used during the DEFCON 8 Capture the Flag (CTF) event [33].

Morin et al. [34] propose a data model for input alert correlation, which allows to aggregate alerts generated by multiple heterogeneous IDSs (e.g. network-based and host-based). The authors state that alert correlation techniques do not take full advantage of the available information about an information system. They identify four main information areas that must be exploited: properties and characteristics of the monitored environment and its vulnerabilities, monitoring systems and events observed.

The model works by correlating input alerts using a similarity function: this function is defined over alerts from the same event (raised by different IDSs), addressing the same vulnerability, belonging to the same TCP/IP connection and based on temporal constraints. No benchmark result is provided to support the system effectiveness.

Lee and Stolfo [35] develop a framework based on data mining techniques, such as sequential patterns mining and episodes rules (see Agrawal and Srikan [36] and Han et al. [37]), to address the problem of improving attack detection while maintaining a low false positive rate. The system works by extracting information from audit traffic and building classification models (specifically designed for certain types of intrusion) using data mining techniques: connection features (i.e. duration, type, protocol) are used to build the *time-based traffic* model, traffic features (i.e. number of connections directed to the same host or same service in a given time frame) constitutes the basis for the *host-based traffic* model while the *content* model collects content features information (i.e. data payload, errors reported by the OS, root access attempts).

The system detects attacks combining the models and comparing them with actual traffic features. Benchmarks have been conducted using the DARPA 1998 data set [38]: detection score for different attack typologies has a minimum value of 65% with a false positive rate always below 0.05%.

### 5.2 Identifying false positives

Pietraszek [15] tackles the problem of reducing false positives by introducing an alert classifier system (**ALAC**, **A**daptive **L**earner for **A**lert **C**lassification) based on machine learning techniques. During the training phase, the system classifies alerts into true positives and false positives, by attaching a label from a fixed set of user-defined labels to the current alert. Then, the system computes an extra parameter (called *classification confidence*) and presents this classification to a human analyst. The analyst's feedback is used to generate training examples, used by the learning algorithm to build and update its classifiers. After the training phase, the classifiers are used to classify new alerts. To ensure the stability of the system over time, a sub-sampling technique is applied: regularly, the system randomly selects $n$ alerts to be forwarded to the analyst instead of processing them autonomously. This approach relies on the analyst's ability to classify alerts properly and on his availability to operate in real-time (otherwise the system will not be updated in time); we believe that these (demanding) requirements can be considered acceptable for a signature-based IDS (where the analyst can easily inspect both the signature and network that triggered the alert), but that it could be difficult to make the same analysis with an anomaly-based system (OAD). Benchmarks conducted over the 1999 DARPA data set [14], using the intrusion detection system Snort [12,13] to generate alerts, show an overall false positives reduction of over 30% (details on single attack classes are not given).

It is worth summarizing the main differences between ALAC and APHRODITE; namely: (a) ALAC does not consider the outgoing traffic, and (b) ALAC relies heavily the expertise and the presence of an analyst (in APHRODITE, all the IT specialist has to do is to set the thresholds). Pietraszek and Tanner [39] further expand the previous work using alert post-processing based on data mining and machine learning techniques.

Julisch [8] presents a semiautomatic approach for identifying true positives based on the idea of *root cause*: an alarm root cause is defined as "the reason for which it occurs". The author observes that in most environments, it is possible to identify a small number of highly predominant (and persistent) root causes. Persistent root causes trigger alarm floods that distract IT specialists from identifying real attacks. The process presented, based on techniques which discover frequently occurring episodes in a given sequence (see Mannila et al. [40,41]), consists of two different steps: the former (called *root cause analysis*) identifies root causes related to a given (large) number of alarms. Then, the latter removes spotted root causes and thereby drastically reduces the future alarm rate.

Benchmarks conducted on a log trace from a commercial NIDS deployed in a real network show a reduction of 87% of root causes. No further details are given about the testing condition, network topology or traffic typology. The work has been further expanded in [42,9] to improve the completeness and accuracy of detecting algorithm.

Qiao and Weixin [28] introduce a NIDS that addresses the problem of reducing false positive combining anomaly-based and signature-based systems and applying a co-stimulation mechanism [43].

The system is composed by two main components: *detectors* and *agent monitors*. The first component is based on a biological immune mechanism (Forrest et al. [44,45]) and it is responsible for detecting attacks: being anomaly-based, it is able to detect zero-day attacks (improving detection rate). Agent monitors are both signature-based and anomaly-based components, which analyze various system parameters and which are responsible for sending a feedback information (the co-stimulation) to the detectors to confirm a possible attack. The agents monitor integrity of sensitive files (integrity monitor), information leakage (confidentiality monitor) and anomaly occupation of resources, e.g. CPU or system memory, (availability monitor). When a detector $d$ raises an alarm, a timer is started: the detector waits for a period of time $\tau$, called *co-stimulation delay*, for the (possible) feedback sent by at least one of the monitor agents. The feedback can also be sent by an IT specialist (e.g. the network administrator), that can label the alert as a real attack. If no feedback is received within the period $\tau$, the alert is considered a false positive. The benchmarks do not allow a full comparison: the authors report only few details about the used data set (private, with artificial attacks introduced by authors themself), and state that all the attacks have been detected without generating false positives.

## 6 Conclusion

In this paper we present APHRODITE, an architecture for reducing false positives in standard NIDS. The core of APHRODITE consists of an Output Anomaly Detector (OAD): when the standard NIDS placed on the input raises an alert, APHRODITE checks if the communication actually raises an anomaly in the output. When this is the case (and in another couple of exceptional situation), the alarm is forwarded to the IT specialist, otherwise it is discarded.

The fact that the OAD is anomaly-based (rather than signature-based) has various advantages: first, the OAD can adapt to the specific network environment/service; secondly it does not require the definition of new signatures to detect anomalous output. Creating and maintaining a set of signatures for the output traffic is labor intensive, as these signatures would heavily depend on the local application, and would have to be updated each time that the application change its output format.

Benchmarks on the DARPA 1999 data set show that APHRODITE determines a reduction of false positives between 50% and 100% in most of the cases, and that it does not introduce any extra false negative. Tests on our private data set show that APHRODITE is still effective also when it is not trained optimally: APHRODITE can be thus used for reducing false positives also in those cases in which the IT professional wants to set it up quickly, without spending time time at cleaning up the data set to carry out an optimal training.

## A  Appendix: APHRODITE pseudo code

In this Section we give a semi-formal description of how APHRODITE works.

DATA TYPE

$l\ =\ length\ of\ the\ longest\ packet\ payload$
$PAYLOAD\ =\ array\ [1..l]\ of\ [0..255]$
$HOMENET\ =\ set\ of\ IP\ addresses$
   /* hosts inside the monitored network */

$HOST\ =\ RECORD\ [$
      $address:\ IP\ address\ \in\ \mathbb{N}$
      $port:\ TCP\ port\ \in\ \mathbb{N}$
   $]$

```
PACKET  =  RECORD [
      source :  HOST
      destination :  HOST
      payload :  PAYLOAD
   ]

ALARM  =  RECORD [
      alarm :
         −∞ if input IDS is signature − based
         value  ∈  Real if input IDS is anomaly − based
      attacker :  HOST (∉  HOMENET)
      victim :  HOST (∈  HOMENET)
      processed :  BOOLEAN /* track a processed alert by the OAD */
      trueIncident :  BOOLEAN /* alarm is marked as an incident */
      counter :  Integer
         /* packets marked as anomalous in a single communication */
   ]
```

DATA STRUCTURE

$\tau \in \mathbb{N}$
   /* Number of packets used for training phase */
$oad \in IDS$
   /* Anomaly-based IDS analyzing outgoing network traffic */
$outThreshold \in Real$
   /* Numeric value used for anomaly detection by OAD */
$magnitudeThreshold \in Real$
   /* Value used to evaluate input alarm magnitude */
$raisedThreshold \in Integer$
   /* Value used to evaluate alarm-raising packets */
$alarms = set\ of\ ALARM$
   /* List of alarms, received from an IDS monitoring incoming traffic */

INIT PHASE
/* IT specialists set outThreshold, magnitudeThreshold and raisedThreshold values */

TRAINING PHASE

INPUT:
   $p :$ PACKET
      /* outgoing network packet */

/* first, train the OAD with $\tau$ samples */
for $t := 1$ to $\tau$

```
    oad.train(p.source.address, p.source.port, p.payload)
end for
```

TESTING PHASE

INPUT:
   $p$ : *PACKET*
      /* outgoing network packet */

OUTPUT:
   *trueIncidents* : *set of ALARM*

```
for each a ∈ alarms do
    if (match_alarm(a, p) = TRUE) then
        /* the function tracks the packet and checks if it belongs to
        a communication marked as anomalous by the input IDS */
        alarm_level := oad.test(p.source.address, p.source.port, p.payload)
        if (alarm_level > outThreshold) then
            a.trueIncident := TRUE
            trueIncidents.add(a)
        end if
        a.processed := TRUE
    end if
end for

/* Here we consider Exception1 */
for each a ∈ alarms do
    if (a.processed = FALSE) then
        a.trueIncident := TRUE
        trueIncidents.add(a)
        a.processed := TRUE
    end if
end for

/* Here we consider Exception2 */
for each a ∈ alarms do
    if (a.alarm > magnitudeThreshold) then
        a.trueIncident := TRUE
        trueIncidents.add(a)
        a.processed := TRUE
    end if
end for

/* Here we consider Exception3 */
for each a ∈ alarms do
```

```
    if (a.counter > raisedThreshold) then
        a.trueIncident := TRUE
        trueIncidents.add(a)
        a.processed := TRUE
    end if
end for

return trueIncidents
```

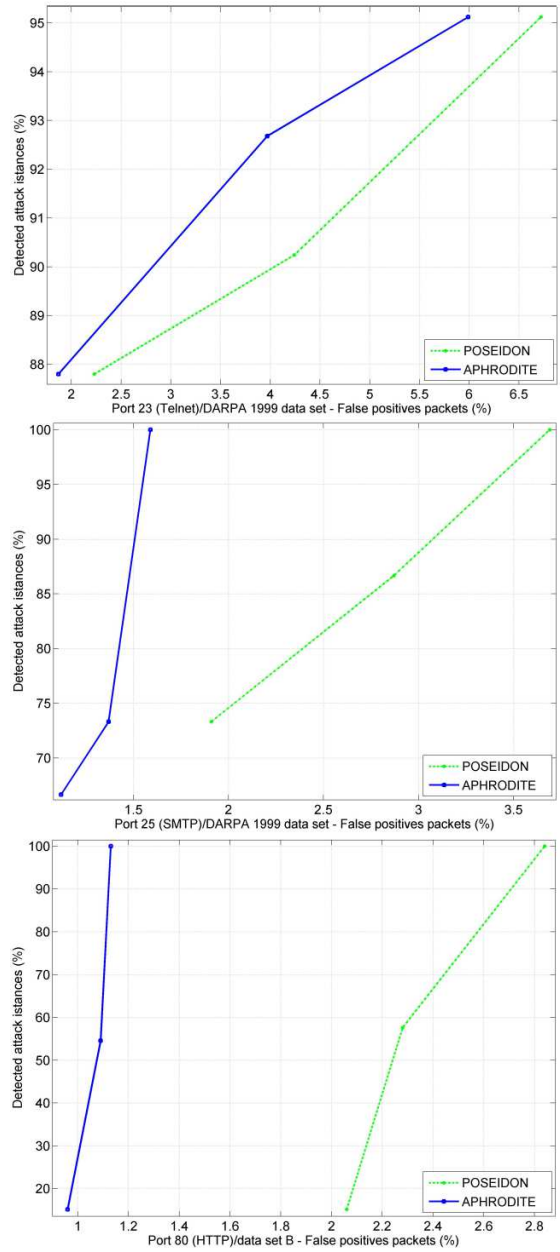

**Fig. 2.** Detection rates for POSEIDON in combination with APHRODITE using DARPA 1999 data set (Telnet and SMTP protocols) and data set B (HTTP protocol): the x-axis and y-axis present false positive rate and detection rate respectively. Is it possible to observe that APHRODITE presents a lower false positive rate than POSEIDON on every benchmarked protocol, considering the same detection rate.